# Sensors and Systems for Monitoring Mental Fatigue: A systematic review

Prabin Sharma[1], Joanna C. Justus[2], Megha Thapa[3] , Govinda R. Poudel[2*]

[1]Department of Computer Science, University of Massachusetts Boston, USA

[2]Mary Mackillop Institute for Health Research, Faculty of Health Sciences, Australian Catholic University, Melbourne, Australia

[3] Nepalese Army Institute of Health Sciences College of Medicine, Kathmandu, Nepal

**Corresponding Author**

Govinda R. Poudel

Mary Mackillop Institute for Health Research

215 Spring Street, Melbourne, Australia

Ph: +6192308368

Govinda.poudel@acu.edu.au

# Abstract

Mental fatigue stands as one of the leading causes of motor vehicle accidents, medical mishaps, decreased workplace productivity, and disengagements among e-learning students. The development of reliable sensors and systems capable of tracking mental fatigue holds the potential to prevent accidents, reduce errors, and enhance workplace productivity. This review offers a critical overview of theoretical models pertaining to mental fatigue, an exploration of key sensor technologies enabling this filed, and a systematic review of recent studies employing biosensors to monitor mental fatigue in humans. We conducted a systematic search and review of recent literature focusing on the detection and monitoring of mental fatigue in humans. The search yielded 57 studies (N=1082), With the majority relying on electroencephalography (EEG)-based sensors for tracking mental fatigue. Interestingly, our findings indicated that EEG- based sensors exhibit moderate to good sensitivity in detecting fatigue, with significant added benefit observed from high-density EEG sensors. These results trigger an essential discussion regarding the integration of wearing EEG devices and ambient sensors for practical real-world monitoring. As we contemplate the future, it's clear that further research and development are needed to make wearable sensors and systems mainstream in semi-autonomous and autonomous industries, ultimately revolutionizing fatigue monitoring practices.

**Introduction**

Mental fatigue is one of the significant contributors to motor vehicle crashes, medical mishaps, and industrial mishaps worldwide[1, 2]. Increased exhaustion links closely with self-reported medical errors among healthcare professionals, impacting up to half of their patients[3, 4]. Fatigue not only hinders productivity in workplaces by inviting distractions and lapses of attention like 'microsleeps' [2, 5], but can also impact team functioning[6]. While safety and risk- management approaches help mitigate the overall harm caused by mental fatigue [7], the urgent need lies in realtime monitoring technologies to reduce the impact on individuals [7]. Recent breakthroughs in sensor technologies and machine learning have ushered in the era of real-time mental fatigue monitoring, especially in safety-critical industries such as mining, aviation, and public transport systems[8]. To ensure widespread adoption, these technologies must be reliable, userfriendly, and secure. This review aims to consolidate the latest advancements in the field of monitoring mental fatigue. Mental fatigue is a unique and personal sensation, like a heavy cloud of weariness, drowsiness, and exhaustion that can make it challenging to carry on with our tasks[9], [10]. This feeling of fatigue can creep in due to various factors, such as disruptions in our sleep patterns, the disarray of our internal body clocks (think jet lag), the relentless strain of monotonous work, psychological and neurological conditions, and outside influences that affect both our minds and bodies[11]. One of the most common culprits behind mental fatigue is a lack of sleep. Extended periods of wakefulness, driven by work commitments or social demands, can intensify our body's urge to rest[12]. This heightened need for sleep can compound mental fatigue, sometimes leading to those fleeting moments of microsleeps[13]. Additionally, disruptions in our circadian rhythms, whether from shift work or jet lag, can trigger a sense of weariness[14]. Shift workers, especially, can find themselves at risk due to the mismatch of their internal body clocks caused by rotating shifts. On the other hand, jet lag, that feeling we get when crossing multiple time zones, can result in daytime tiredness, struggles to stay alert, and a general sense of not feeling quite right [11, 15]. Mental fatigue that arises from the extensive or prolonged use of our mental or physical abilities is often referred to as "use-dependent" fatigue. Tasks that are monotonous and demand continuous attention, such as long highway drives, equipment monitoring, or operating highly automated aircraft, can deplete our mental resources and give rise to this kind of mental fatigue [2, 16]. In controlled studies, lapses in concentration have been observed in as little as 20 minutes while performing repetitive tracking tasks [17]. Even during monotonous driving, errors can surface within the first 10 minutes, occasionally leading to brief episodes of microsleeps, even when a person is well-rested [18, 19]. Furthermore, external environmental factors play a significant role. Elements like the time of day, temperature, humidity, altitude, noise levels, illumination, and vibration can all contribute to inducing fatigue [11]. Excessive noise and vibrations in the workplace have been positively linked to heightened fatigue levels [20, 21]. Notably, fatigue is also a common symptom in various neurological disorders and progressive neurodegenerative conditions [22]. In these cases, fatigue can result from the direct impact of underlying neurodegeneration or secondary to disruptions in sleep-wake patterns [23].

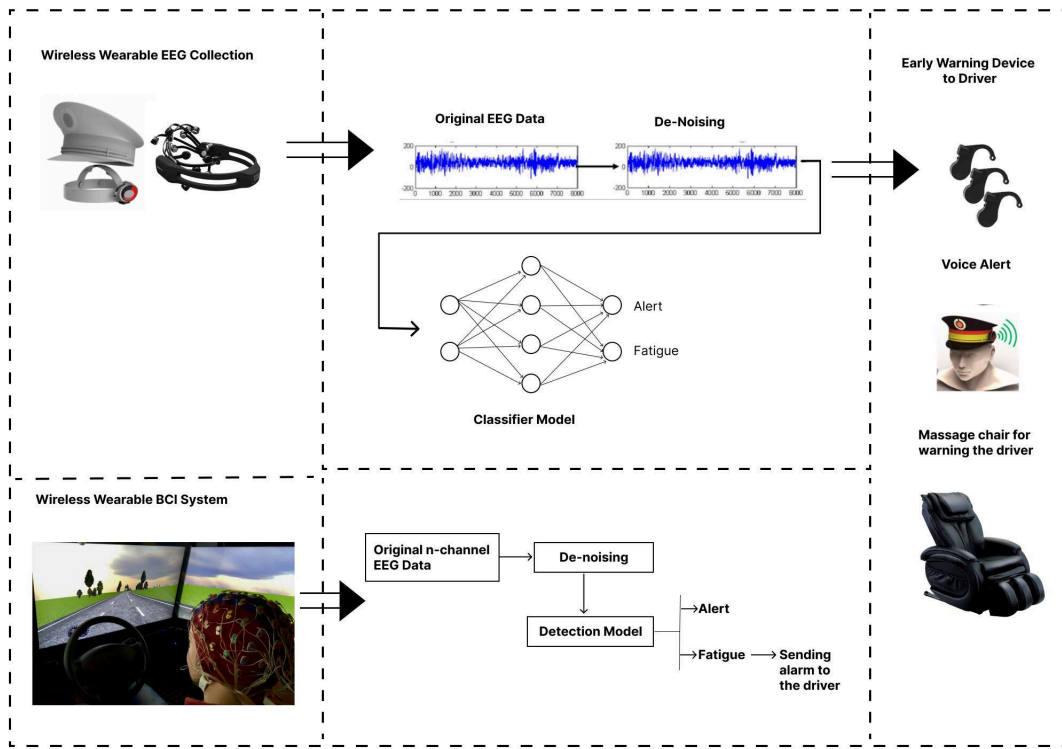

**Figure 1:** A flow diagram of the system which uses EEG-based sensors for monitoring mental fatigue.

Mental fatigue can reveal itself through lapses in attention, and these lapses can take various forms. Some are subtle, like delayed responses, while others represent a complete breakdown in our ability to respond effectively. Some of these lapses come with behavioral indicators, such as moments of eye-closure, a loss of muscle tone, or even a sense of drowsiness [24]. Moreover, changes in the brain's electrical activity, as seen in EEG characteristics, have been linked to these lapses [25, 26]. Exciting recent discoveries have shed light on the different brain processes behind lapses during states of reduced alertness, like sleep deprivation, and lapses during well-rested periods [27, 28]. During continuous tasks like visuomotor tracking, lapses in attention can manifest as a flat or disjointed response, often accompanied by a drop in response speed. This phenomenon, known as 'lapses of responsiveness' (LOR), is particularly intriguing due to the continuous nature of these tasks, which allows us to pinpoint even the slightest fluctuations in sensory-motor performance resulting from lowered arousal or lapses in focus[29]. For instance, in one study, Peiris et al. (2006) utilized a visuomotor tracking task to identify instances of complete failure to respond, even in individuals who weren't sleep-deprived. Huang et al. (2008) also observed similar instances of complete non-responsiveness during a compensatory tracking task, which they classified as lapses of responsiveness [30]. Measuring mental fatigue involves a range of approaches, neatly divided into three categories: performance measures, behavioral measures, and physiological measures. Performance measures keep a watchful eye on changes in a person's responsiveness, tracking factors like response times, lane deviations, and even how tightly they grip the steering wheel [29]. Behavioral approaches take advantage of eye-tracking technology to gauge levels of mental fatigue [30]. Meanwhile, physiological measures delve into the realm of brain activity, utilizing EEG to detect neural signals via electrodes on the scalp [29][31]. Of these methods, behavioural and physiological measures emerge as the most promising options for real-time and non-intrusive monitoring of mental fatigue. Our systematic review aims to consolidate the latest advancements in physiological sensors and systems designed for precise mental fatigue monitoring.

## Method

**Search Strategy**

This comprehensive review adhered to the 2009 PRISMA statement [32]. Extensive searches were meticulously carried out using the APA PsycINFO and Medline Complete electronic databases, without imposing any date restrictions. The PICO Model served as our guiding framework, especially in the definition of search terms for both comparator and outcome components. These search terms were thoughtfully combined using Boolean operators 'AND' and 'OR'.

We carefully searched through titles and abstracts to learn more about fatigue and prediction methods. Our search terms included 'microsleep,' 'micro sleep,' 'micro-sleep,' 'fatigue,' and 'drows*' combined with 'artificial intelligence,' 'machine learning,' 'deep learning,' 'classification,' and 'regression.' Additionally, subject headings such as 'Sleepiness,' 'Fatigue,' 'Artificial Intelligence,' 'Machine Learning,' 'Deep Learning' (MeSH only), and 'Classification' (MeSH only) were employed to ensure a comprehensive exploration of the subject matter.

We broadened our search scope by investigating terms related to EEG and facial features within titles and abstracts, using 'OR' as separators to encompass a wider range of search results. These terms, such as 'EEG,' 'Electroencephalogram,' 'eye closure,' 'eye video,' 'focal,' 'perclos,' and 'pupil' were thoughtfully combined with 'OR' to ensure comprehensive results. To refine our search, we employed the subject heading 'Electroencephalography' for these terms, which are noted in the Appendix for reference.

**Eligibility Criteria**

We exclusively considered recent publications, all within the last decade, for our systematic review. Publications in languages other than English were omitted. Our initial screening, based on specific criteria, was twofold. First, we assessed titles and abstracts, excluding studies with (a) clinical samples, (b) animal samples, (c) theses or dissertations, or (d) book chapters. Subsequently, we performed full-text screening, excluding studies that met any of these criteria: (a) included clinical samples, (b) didn't aim to detect fatigue, (c) were non-empirical, (d) were nonexperimental, (e) didn't provide accuracy percentages, or (f) omitted the method of prediction. Clinical samples were excluded because our research's focus was on detecting fatigue in healthy individuals. Clinical samples included participants with psychopathologies [33] or sleep disorders [34]. Among the studies excluded during full-text screening (n= 25), many didn't aim to assess fatigue detection. Instead, they concentrated on prediction methods with different objectives, such as muscle fatigue detection [35] or sleep stage classification [36].

Studies were classified as non-experimental if they didn't involve participants in tasks like driving simulations or cognitive tasks [37], [10]. Similarly, studies were considered non-empirical if they consisted of theoretical discussions, such as presenting a novel model [7] or summarizing existing research [38]. Furthermore, studies were eliminated if they lacked accuracy percentage reporting. Some of these studies reported accuracy based on task errors [39] or reaction times [40], while others utilized AUC and ROC ([41], [42]). One study was excluded because it didn't disclose its prediction method [43]. Figure 2 illustrates the PRISMA Flowchart, detailing our selection process.

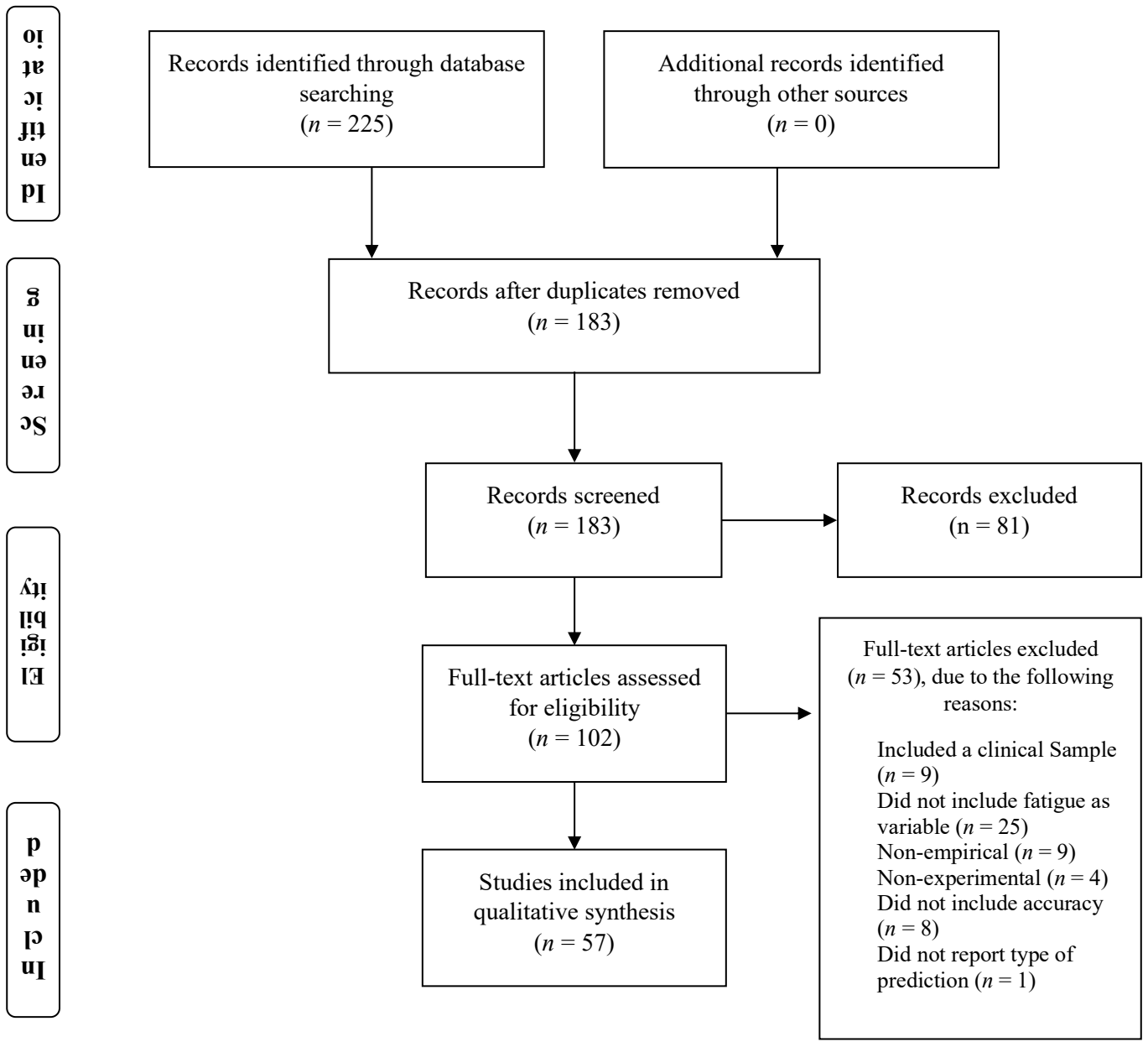

**Figure 2:** *PRISMA flowchart for selection process*

**Data Extraction**

After full-text screening, information from each publication was extracted. Descriptive statistics such as publication year, country of origin, sample size, and age characteristics were focused on. Types of measures used, experimental stimulus, method of prediction, and resulting accuracy were also extracted from each study.

All measures employed were collected, with some studies incorporating one measure, while others included two or more. When EEG was utilized, the number of channels was documented. In terms of the experimental stimulus, all studies relied upon one of four options: (a) simulator, (b) real-life task, (c) cognitive task, or (d) another form of stimulus. The method of prediction was systematically categorized into classification, neural network, and linear regression.

The resulting accuracy was recorded as a percentage. Depending on the data reported, either the average accuracy or the highest accuracy was extracted. For studies employing multiple measures or prediction methods, the average accuracy for each was calculated [44], [45]. In a specific study comparing driving simulator and cognitive tasks, two accuracy percentages were recorded [46]. Additionally, two studies [47], [29] presented multiple averages due to their utilization of various methods, with the highest average being selected.

## Results

### Study characteristics

Our initial search yielded a total of 225 publications. Following the removal of duplicates and a meticulous screening process to identify relevant papers, we ultimately included 57 studies in this review. For a comprehensive list of these papers and the extracted information, please refer to supplementary Table 1. The mean sample size was 19, with a standard deviation of 13. Among the 57 studies, the predominant experimental paradigm employed was the use of a driving simulator (n=42). Other paradigms encompassed the use of flying simulators, real driving or flying scenarios, cognitive tasks, daytime naps, or maintenance of wakefulness tests. The driving simulator frequently utilized was the Divided Attention Steering Simulator (DASS), as seen in various studies [48]. This simulator was equipped with typical car controls such as brake, accelerator, and steering wheel [49]. It featured a substantial screen displaying dashboard elements like the odometer, as well as other road-related stimuli like other vehicles. To immerse participants further, engine and traffic noises were integrated to replicate a realistic driving environment [50]. Participants were given the task of maintaining their vehicle in the center of the road for as long as possible [48]. The duration of simulator sessions varied, spanning from 30 minutes to 6 hours. When signs of fatigue were detected, the simulator automatically ceased the exercise. In some instances, participants received a practice trial before the actual simulation. It was a requirement for participants to possess a valid driver's license, and many studies explicitly stated that participants should abstain from caffeine and alcohol while ensuring they had at least 7 hours of sleep the night before the experiment. The specific driving route varied among studies, with some simulating scenarios like motorway driving with a speed limit of 100 km/h. The choice of route depended on the desired mental effort level for the participants [51]. Additionally, certain simulators incorporated particular driving conditions, such as sunny or snowy weather, and routes that included mountainous terrain or countryside settings [52].

**Sensor for monitoring fatigue**

In all the studies, EEG served as the primary measurement tool; however, there was variation in the number of EEG channels utilized (Mean = 26, Standard Deviation = 19). Additional measures were incorporated in some studies, including EOG, ECG, EMG, psychometric data, or video recordings. Furthermore, in certain cases, the inclusion of respiration signals, PPG, and GSR data was observed. The total count of EEG channels employed in these studies was categorized into six groups: 1–9, 10–19, 20–29, 30–39, 40–49, and 50 or more, as detailed in Figure 1, where the mean accuracy based on EEG channel grouping is presented. In all the studies we looked at, they achieved accuracy rates higher than 59.7%, and the best result reached an impressive 99.1% (M = 84.8, SD = 10.2). with the highest accuracy reaching an impressive 99.1% (M = 84.8, SD = 10.2). An analysis using a one-way ANOVA demonstrated that the differences in mean accuracy for each EEG channel grouping were not found to be statistically significant, as denoted by $F(5, 43) = 1.66$, $p = .17$.

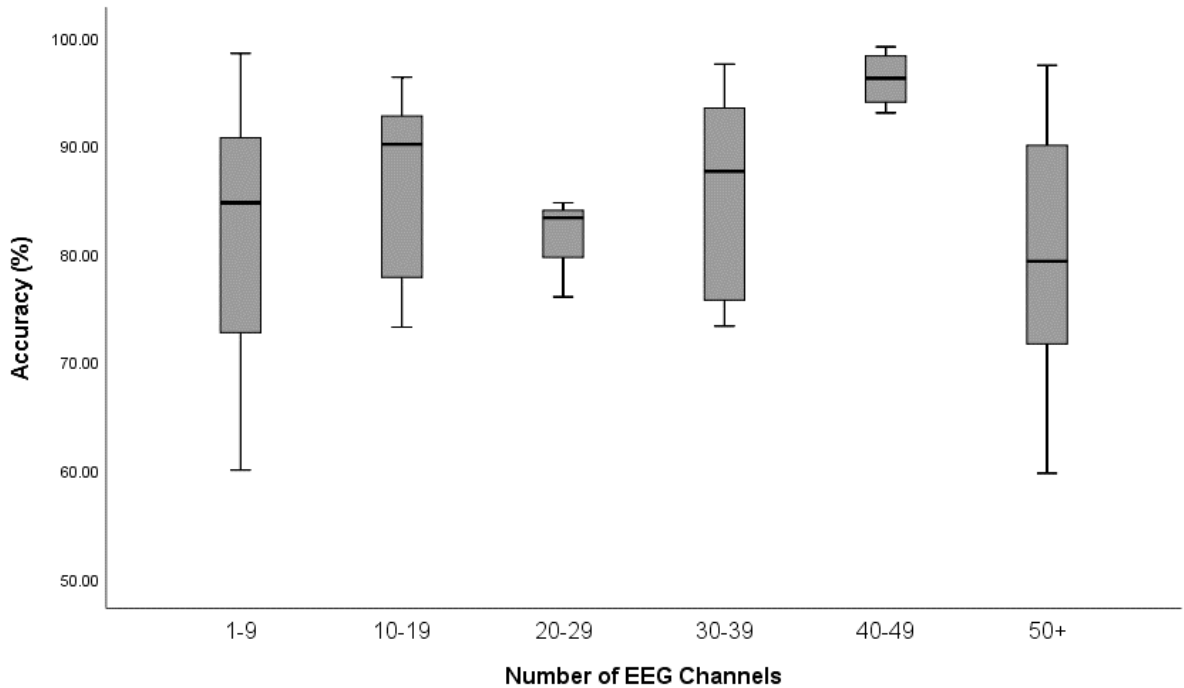

Figure 3: Accuracy (%) of prediction of mental fatigue grouped according to the number of EEG electrodes used in the study.

**Method of Prediction**

Three primary prediction approaches were employed by the studies: 1) classification, 2) neural network, or 3) linear regression. The mean accuracy based on the method of prediction is depicted in Figure 2. Notably, a higher mean accuracy is observed for both classification and neural network methods compared to linear regression. To investigate these differences further, a one-way analysis of variance (ANOVA) was conducted. The results revealed that the variations in mean accuracy for these prediction methods were not found to be statistically significant, as indicated by $F(2, 49) = 1.00$, $p = .38$. The means and standard deviations for each method of prediction are detailed in Table 1. It should be noted that there are unequal sample sizes among the different prediction methods, with a particularly low sample size observed for linear regression.

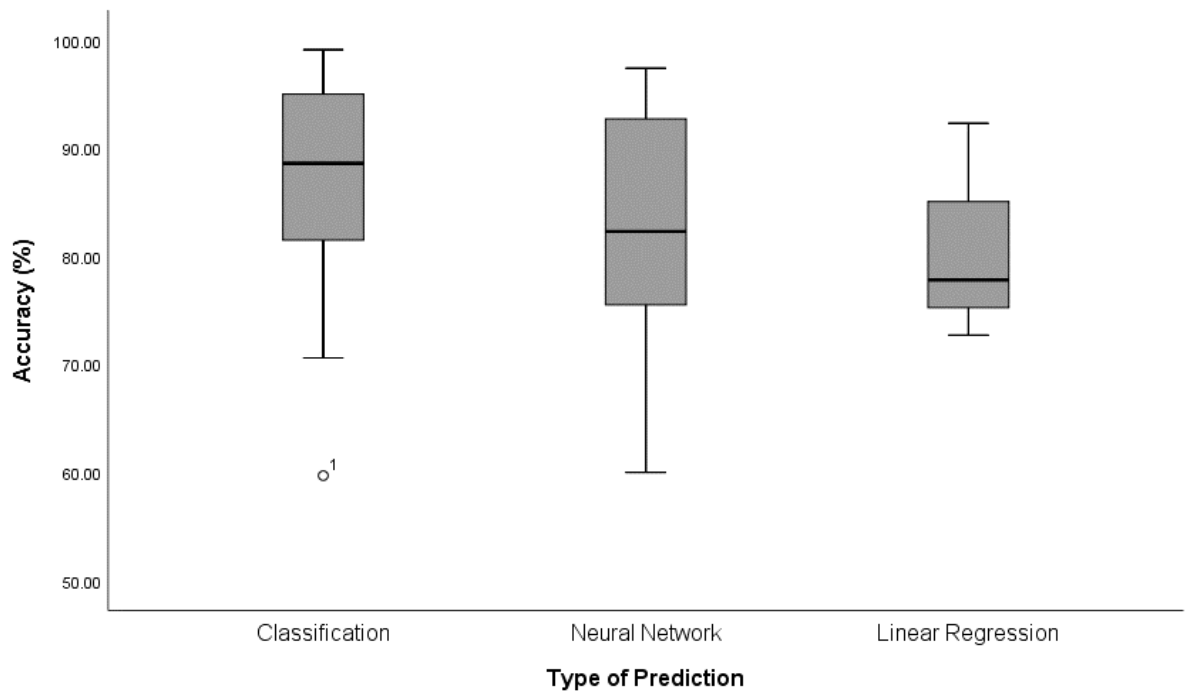

Figure 4: Accuracy (%) of prediction of mental fatigue grouped according to the type of method used for predicting mental fatigue.

**Discussion**

This systematic review presents a comprehensive synthesis of research in the field of mental fatigue detection. The findings from our analysis of 57 studies indicate that there are no significant differences in accuracy based on the number of EEG channels used.

The majority of these studies relied on EEG-based sensors for monitoring mental fatigue. Surprisingly, the quantity of electrodes used in predicting mental fatigue does not appear to have a significant impact on the accuracy of these predictions. This discovery challenges the traditional notion that more electrodes lead to better prediction accuracy.

This finding carries practical implications, suggesting that using fewer electrodes may reduce the cost and complexity of prediction systems without compromising accuracy. It's possible that this result is due to the nature of brain activity associated with mental fatigue, which is distributed across different regions of the brain. Therefore, increasing the number of electrodes may not necessarily capture more relevant information. Additionally, using more electrodes may introduce more noise, potentially reducing prediction accuracy. However, further research is required to confirm this hypothesis.

Previous research has indicated that the placement and configuration of EEG electrodes may play a pivotal role in capturing precise and dependable signals associated with mental fatigue [25, 26]. The choice of electrode locations, including frontal, central, and parietal regions, can profoundly influence the specificity and sensitivity of fatigue detection [25, 26]. Furthermore, the number of electrodes employed and their spatial distribution can impact the resolution and coverage of brain activity monitoring. Factors such as electrode size, material, and impedance can also influence signal quality and the electrode-skin interface [27, 28].

The careful selection and preparation of electrodes are vital to minimizing artifacts and ensuring the collection of high-quality data. Moreover, the ongoing advancements in wearable and dry electrode technologies hold the potential to enhance the convenience, comfort, and accessibility of EEG-based fatigue detection. In this review, it was observed that the experimental stimuli used varied between studies. For example, driving or flight simulators were used in some studies, while real-life driving or cognitive tasks were employed in others. This diversity in participant experiences can lead to differing outcomes due to variations in mental and physical demands. It should be acknowledged that when applying findings to real-world driving and related scenarios, consideration should be given to this diversity. Notably, findings from flight simulators may not be applicable to the broader population of healthy drivers.

While driving simulators were predominantly used in the systematic review, there was significant variation in the simulations themselves, including motorway routes, countryside settings, and different road and weather conditions. This variability is expected given the diverse driving conditions in various contexts, but it must be considered when generalizing fatigue detection research.

Real-life scenarios were only utilized in two studies, prompting a suggestion from fatigue detection researchers that a preference should be given to real-life situations over simulations. This is because simulations might not faithfully replicate all facets of actual driving, as acknowledged in reference [53].

Considering the focus of our review on healthy individuals in driving contexts, it is advisable that future fatigue detection research concentrate exclusively on demographics prone to fatigue-related driving scenarios, such as truck drivers and taxi drivers. This approach allows populations more likely to encounter motor vehicle accidents or similar situations to be targeted by researchers. Moreover, it is recommended that future studies use real-life driving routes that mirror the typical driving conditions of the specific population, including typical road and weather conditions. Given the prevailing emphasis on fatigue while driving, related scenarios like pilot fatigue should be explored in future research. Additionally, fatigue detection research could be of benefit to the medical field, particularly in professions involving monotonous work.

For a fatigue detection system to achieve ergonomic user-friendliness, the necessity arises for it to be both portable and easily wearable by drivers. Recent technological advancements have made it feasible to implement even intricate classification algorithms on portable devices, such as mobile phones. Consequently, the prospect emerges for future devices to employ advanced algorithms rooted in neural networks for fatigue prediction. As exemplified in the study conducted by Liu, Chiang et al. (2013), portable devices can leverage software applications across platforms like Android, iPhone, and DotNET.

When it comes to predicting fatigue using EEG (brainwave data), it might be smarter to use fewer EEG channels, as suggested by reference [48], as this entails less setup time and shows no noticeable reduction in accuracy. Given the focus of our review on EEGs, it is recommended that future reviews explore the synergy of various measures in fatigue detection.

In summary, this systematic review provides a synthesis of existing research on fatigue detection. The results from the analysis of 46 studies suggest that there are no significant differences in accuracy based on the number of EEG channels used or the type of prediction method employed. These findings point towards the investigation of linear regression methods and the exploration of combinations of fatigue detection measures to better understand the variables contributing to higher accuracy. Future research is encouraged to focus on populations more likely to experience fatigue while driving and to utilize real-life scenarios instead of simulations. Additionally, consideration should be given to the use of comfortable EEG caps and portable devices from an ergonomic perspective.

# References

1. Dorrian, J.; Chapman, J.; Bowditch, L.; Balfe, N.; Naweed, A., A survey of train driver schedules, sleep, wellbeing, and driving performance in Australia and New Zealand. *Sci Rep-Uk* **2022,** 12, (1).
2. Caldwell, J. A.; Caldwell, J. L.; Thompson, L. A.; Lieberman, H. R., Fatigue and its management in the workplace. *Neurosci Biobehav Rev* **2019,** 96, 272–289.
3. West, C. P.; Tan, A. D.; Habermann, T. M.; Sloan, J. A.; Shanafelt, T. D., Association of resident fatigue and distress with perceived medical errors. *JAMA* **2009,** 302, (12), 1294–1300.
4. Huizinga, C. R. H.; Tummers, F.; Mheen, P. J. M.-v.; Cohen, A. F.; Bogt, K. E. A., A review of current approaches for evaluating impaired performance in around-the-clock medical professionals. *Sleep Med Rev* **2019,** 46, 97–107.
5. Poudel, G. R.; Hawes, S.; Innes, C. R. H.; Parsons, N.; Drummond, S. P. A.; Caeyensberghs, K.; Jones, R. D., RoWDI: rolling window detection of sleep intrusions in the awake brain using fMRI. *J Neural Eng* **2021,** 18, (5).
6. Banks, S.; Landon, L. B.; Dorrian, J.; Waggoner, L. B.; Centofanti, S. A.; Roma, P. G.; Van Dongen, H. P. A., Effects of fatigue on teams and their role in 24/7 operations. *Sleep Med Rev* **2019,** 48.
7. Zhang, Z.; Luo, D.; Rasim, Y.; Li, Y.; Meng, G.; Xu, J.; Wang, C., A Vehicle Active Safety Model: Vehicle Speed Control Based on Driver Vigilance Detection Using Wearable EEG and Sparse Representation. *Sensors* **2016,** 16, (2), 242.
8. Martins, N. A. R.; Annaheim, S.; Spengler, C. M.; Rossi, R. M., Fatigue Monitoring Through Wearables: A State-of-the-Art Review. *Front Physiol* **2021,** 12.
9. Brown, I. D., Driver fatigue. *Hum Factors* **1994,** 36, (2), 298–314.
10. Komarov, O.; Ko, L. W.; Jung, T. P., Associations Among Emotional State, Sleep Quality, and Resting-State EEG Spectra: A Longitudinal Study in Graduate Students. *IEEE transactions on neural systems and rehabilitation engineering : a publication of the IEEE Engineering in Medicine and Biology Society* **2020,** 28, (4), 795–804.
11. Sanjaya, K. H.; Lee, S.; Katsuura, T., Review on the application of physiological and biomechanical measurement methods in driving fatigue detection. **2016,** 7, 14.
12. Doran, S. M.; Van Dongen, H. P. A.; Dinges, D. F., Sustained attention performance during sleep deprivation: Evidence of state instability. *Arch Ital Biol* **2001,** 139, (3), 253-267.
13. Poudel, G. R.; Innes, C. R. H.; Jones, R. D., Temporal evolution of neural activity and connectivity during microsleeps when rested and following sleep restriction. *Neuroimage* **2018,** 174, 263-273.
14. Sorengaard, T. A.; Saksvik-Lehouillier, I.; Langvik, E., Longitudinal and cross-sectional examination of the relationship between personality and fatigue among shift workers. *Cogent Psychol* **2019,** 6, (1).
15. Sanjaya, K. H.; Sya'Bana, Y. M. K.; Hutchinson, S.; Diels, C., Preliminary investigation of sleep-related driving fatigue experiment in Indonesia. *Journal of Mechatronics, Electrical Power, and Vehicular Technology* **2018,** 9, (1), 8-16.
16. Guo, M.; Li, S.; Wang, L.; Chai, M.; Chen, F.; Wei, Y., Research on the Relationship between Reaction Ability and Mental State for Online Assessment of Driving Fatigue. *International journal of environmental research and public health* **2016,** 13, (12).
17. Poudel, G. R.; Innes, C. R. H.; Bones, P. J.; Watts, R.; Jones, R. D., Losing the Struggle to Stay Awake: Divergent Thalamic and Cortical Activity During Microsleeps. *Hum Brain Mapp* **2014,** 35, (1), 257-269.
18. Ting, P. H.; Hwang, J. R.; Doong, J. L.; Jeng, M. C., Driver fatigue and highway driving: A simulator study. *Physiol Behav* **2008,** 94, (3), 448-453.
19. Peiris, M. T.; Jones, R. D.; Davidson, P. R.; Carroll, G. J.; Bones, P. J., Frequent lapses of responsiveness during an extended visuomotor tracking task in non-sleep-deprived subjects. *Journal of sleep research* **2006,** 15, (3), 291–300.

20. Lee, S.; Katsuura, T.; Shimomura, Y.; Liu, X.; Konno, F.; Onishi, M.; Tada, M.; Kotegawa, K., Effects of Active Noise Control on Physiological Functions. *Journal of the Human-Environment System* **2009,** 12, 49–54.
21. Taylor, J. L.; Amann, M.; Duchateau, J.; Meeusen, R.; Rice, C. L., Neural Contributions to Muscle Fatigue: From the Brain to the Muscle and Back Again. *Medicine and science in sports and exercise* **2016,** 48, (11), 2294–2306.
22. Manjaly, Z. M.; Harrison, N. A.; Critchley, H. D.; Do, C. T.; Stefanics, G.; Wenderoth, N.; Lutterotti, A.; Muller, A.; Stephan, K. E., Pathophysiological and cognitive mechanisms of fatigue in multiple sclerosis. *J Neurol Neurosurg Psychiatry* **2019,** 90, (6), 642-651.
23. Baker, C. R.; Dominguez, D. J.; Stout, J. C.; Gabery, S.; Churchyard, A.; Chua, P.; Egan, G. F.; Petersen, A.; Georgiou-Karistianis, N.; Poudel, G. R., Subjective sleep problems in Huntington's disease: A pilot investigation of the relationship to brain structure, neurocognitive, and neuropsychiatric function. *J Neurol Sci* **2016,** 364, 148-53.
24. Peiris, M.; Jones, R.; Davidson, P.; Carroll, G.; Bones, P., Frequent lapses of responsiveness during an extended visuomotor tracking task in non-sleep-deprived subjects. *Journal of Sleep Research* **2006,** 15, 291-300.
25. Williams, H.; Granda, A.; Jones, R.; Lubin, A.; Armington, J., EEG frequency and finger pulse volume as predictors of reaction time during sleep loss. *Electroencephalography and Clinical Neurophysiology* **1961,** 14, 64-70.
26. Makeig, S.; Inlow, M., Lapse in alertness: coherence of fluctuations in performance and EEG spectrum. *Electroencephalography and Clinical Neurophysiology* **1993,** 86, (1), 23-35.
27. Weissman, D.; Roberts, K.; Visscher, K.; Woldorff, M., The neural bases of momentary lapses in attention. *Nature Neuroscience* **2006,** 9, 971-978.
28. Chee, M.; Tan, J.; Zheng, H.; Parimal, S.; Weissman, D.; Zagorodnov, V.; Dinges, D., Lapsing during sleep deprivation is associated with distributed changes in brain activation. *Journal of Neuroscience* **2008,** 28, (21), 5519.
29. Zeng, H.; Yang, C.; Zhang, H.; Wu, Z.; Zhang, J.; Dai, G.; Babiloni, F.; Kong, W., A LightGBM-Based EEG Analysis Method for Driver Mental States Classification. *Computational intelligence and neuroscience* **2019**, 3761203.
30. Zhang, X.; Li, J.; Liu, Y.; Zhang, Z.; Wang, Z.; Luo, D.; Zhou, X.; Zhu, M.; Salman, W.; Hu, G.; Wang, C., Design of a Fatigue Detection System for High-Speed Trains Based on Driver Vigilance Using a Wireless Wearable EEG. **2017**.
31. Lal, S. K. L.; Craig, A., Driver fatigue: Electroencephalography and psychological assessment. *Psychophysiol* **2002,** 39, (3), 313–321.
32. Moher, D.; Liberati, A.; Tetzlaff, J.; Altman, D. G.; Group, P., Preferred Reporting Items for Systematic Reviews and Meta-Analyses: The PRISMA Statement. *Annals of Internal Medicine* **2009,** 151, (4), 264–269.
33. Chelaru, M. I.; Slater, J. D., A Continuous Clustering Algorithm for Detection of Local Sleep in Humans. *The Journal of neuropsychiatry and clinical neurosciences* **2019,** 31, (4), 353–360.
34. Skorucak, J.; Hertig-Godeschalk, A.; Schreier, D. R.; Malafeev, A.; Mathis, J.; Achermann, P., Automatic detection of microsleep episodes with feature-based machine learning. *Sleep* **2020,** 43, (1).
35. Tryon, J.; Friedman, E.; Trejos, A. L., Performance Evaluation of EEG/EMG Fusion Methods for Motion Classification. In *IEEE ... International Conference on Rehabilitation Robotics : [proceedings*, 2019; pp 971–976.
36. Michielli, N.; Acharya, U. R.; Molinari, F., Cascaded LSTM recurrent neural network for automated sleep stage classification using single-channel EEG signals. *Computers in biology and medicine* **2019,** 106, 71–81.
37. Punsawad, Y.; Wongsawat, Y., A multi-command SSVEP-based BCI system based on single flickering frequency half-field steady-state visual stimulation. *Medical & biological engineering & computing* **2017,** 55, (6), 965–977.
38. Sander, C.; Hensch, T.; Wittekind, D. A.; Böttger, D.; Hegerl, U., Assessment of Wakefulness and Brain Arousal Regulation in Psychiatric Research. *Neuropsychobiology* **2015,** 72, (3-4), 195–205.

39. Ndaro, N. Z.; Wang, S. Y., Effects of Fatigue Based on Electroencephalography Signal during Laparoscopic Surgical Simulation. *Minimally invasive surgery* **2018**, 2389158.
40. Tian, S.; Wang, Y.; Dong, G.; Pei, W.; Chen, H., Mental Fatigue Estimation Using EEG in a Vigilance Task and Resting States. In *Conference proceedings : ... Annual International Conference of the IEEE Engineering in Medicine and Biology Society. IEEE Engineering in Medicine and Biology Society. Annual Conference 2018: 1980-1983*, 2018.
41. Hajinoroozi, M.; Jianqiu, Z.; Yufei, H., Prediction of fatigue-related driver performance from EEG data by deep Riemannian model. In *Conference proceedings : ... Annual International Conference of the IEEE Engineering in Medicine and Biology Society. IEEE Engineering in Medicine and Biology Society. Annual Conference 2017*, 2017; pp 4167–4170.
42. Shoorangiz, R.; Buriro, A. B.; Weddell, S. J.; Jones, R. D., Detection and Prediction of Microsleeps from EEG using Spatio-Temporal Patterns. In *Conference proceedings : ... Annual International Conference of the IEEE Engineering in Medicine and Biology Society. IEEE Engineering in Medicine and Biology Society. Annual Conference 2019*, 2019; pp 522–525.
43. Kleifges, K.; Bigdely-Shamlo, N.; Kerick, S. E.; Robbins, K. A., BLINKER: Automated Extraction of Ocular Indices from EEG Enabling Large-Scale Analysis. *Frontiers in neuroscience* **2017,** 11, 12.
44. Taeho, H.; Miyoung, K.; Seunghyeok, H.; Suk, P. K., Driver drowsiness detection using the in-ear EEG. In *Conference proceedings : ... Annual International Conference of the IEEE Engineering in Medicine and Biology Society. IEEE Engineering in Medicine and Biology Society. Annual Conference 2016*, 2016; pp 4646–4649.
45. Hernández, L. G.; Mozos, O. M.; Ferrández, J. M.; Antelis, J. M., EEG-Based Detection of Braking Intention Under Different Car Driving Conditions. *Frontiers in neuroinformatics* **2018,** 12, 29.
46. Dimitrakopoulos, G. N.; Kakkos, I.; Dai, Z.; Wang, H.; Sgarbas, K.; Thakor, N.; Bezerianos, A.; Sun, Y., Functional Connectivity Analysis of Mental Fatigue Reveals Different Network Topological Alterations Between Driving and Vigilance Tasks. *IEEE transactions on neural systems and rehabilitation engineering : a publication of the IEEE Engineering in Medicine and Biology Society* **2018,** 26, (4), 740–749.
47. Zeng, H.; Yang, C.; Dai, G.; Qin, F.; Zhang, J.; Kong, W., EEG classification of driver mental states by deep learning. *Cognitive neurodynamics* **2018,** 12, (6), 597–606.
48. Chai, R.; Tran, Y.; Craig, A.; Ling, S. H.; Nguyen, H. T., Enhancing accuracy of mental fatigue classification using advanced computational intelligence in an electroencephalography system. In *Conference proceedings : ... Annual International Conference of the IEEE Engineering in Medicine and Biology Society. IEEE Engineering in Medicine and Biology Society. Annual Conference*, 2014; pp 1338–1341.
49. Cai, Q.; Gao, Z. K.; Yang, Y. X.; Dang, W. D.; Grebogi, C., Multiplex Limited Penetrable Horizontal Visibility Graph from EEG Signals for Driver Fatigue Detection. *International journal of neural systems* **2019,** 29, (5), 1850057.
50. Chen, J.; Wang, H.; Hua, C., Assessment of driver drowsiness using electroencephalogram signals based on multiple functional brain networks. *International journal of psychophysiology : official journal of the International Organization of Psychophysiology* **2018,** 133, 120–130.
51. Dimitrakopoulos, G. N.; Kakkos, I.; Thakor, N. V.; Bezerianos, A.; Yu, S., A mental fatigue index based on regression using mulitband EEG features with application in simulated driving. In *Conference proceedings : ... Annual International Conference of the IEEE Engineering in Medicine and Biology Society. IEEE Engineering in Medicine and Biology Society. Annual Conference 2017*, 2017; pp 3220–3223.
52. Hu, J.; Min, J., Automated detection of driver fatigue based on EEG signals using gradient boosting decision tree model. *Cognitive neurodynamics* **2018,** 12, (4), 431–440.
53. Akerstedt, T.; Ingre, M.; Kecklund, G.; Anund, A.; Sandberg, D.; Wahde, M.; Philip, P.; Kronberg, P., Reaction of sleepiness indicators to partial sleep deprivation, time of day and time on task in a driving simulator--the DROWSI project. *Journal of sleep research* **2010,** 19, (2), 298–309.

54. Sun, Y.; Lim, J.; Meng, J.; Kwok, K.; Thakor, N.; Bezerianos, A., Discriminative analysis of brain functional connectivity patterns for mental fatigue classification. *Annals of biomedical engineering* **2014,** 42, (10), 2084–2094.